\newcommand{\CLASSINPUTtoptextmargin}{19mm}%
\newcommand{\CLASSINPUTbottomtextmargin}{43mm}%
\newcommand{\CLASSINPUTinnersidemargin}{12.9mm}%
\newcommand{\CLASSINPUToutersidemargin}{12.9mm}%
\documentclass[conference,10pt,a4paper]{IEEEtran}
\usepackage{amsmath}
\usepackage{times}
\usepackage{graphicx}
\usepackage{multirow}
\usepackage[none]{hyphenat}
\usepackage{float}
\usepackage{subfig}

\usepackage[plain]{fancyref}
\makeatletter

\def\@maketitle{\newpage
\bgroup\par\addvspace{0.5\baselineskip}\centering%
\ifCLASSOPTIONtechnote
   {\bfseries\large\@IEEEcompsoconly{\sffamily}\@title\par}\vskip 1.3em{\lineskip .5em\@IEEEcompsoconly{\sffamily}\@author
   \@IEEEspecialpapernotice\par{\@IEEEcompsoconly{\vskip 1.5em\relax
   \@IEEEtitleabstractindextextbox{\@IEEEtitleabstractindextext}\par
   \hfill\@IEEEcompsocdiamondline\hfill\hbox{}\par}}}\relax
\else
   \vskip0.2em{\EuMWtitlesize\ifCLASSOPTIONtransmag\bfseries\LARGE\fi\@IEEEcompsoconly{\sffamily}\@IEEEcompsocconfonly{\normalfont\normalsize\vskip 2\@IEEEnormalsizeunitybaselineskip
   \bfseries\Large}\@title\par}\vskip1.0em\par
   \ifCLASSOPTIONconference%
      {\@IEEEspecialpapernotice\mbox{}\vskip\@IEEEauthorblockconfadjspace%
       \mbox{}\hfill\begin{@IEEEauthorhalign}\@author\end{@IEEEauthorhalign}\hfill\mbox{}\par}\relax
   \else
      \ifCLASSOPTIONpeerreviewca
         {\@IEEEcompsoconly{\sffamily}\@IEEEspecialpapernotice\mbox{}\vskip\@IEEEauthorblockconfadjspace%
          \mbox{}\hfill\begin{@IEEEauthorhalign}\@author\end{@IEEEauthorhalign}\hfill\mbox{}\par
          {\@IEEEcompsoconly{\vskip 1.5em\relax
           \@IEEEtitleabstractindextextbox{\@IEEEtitleabstractindextext}\par\hfill
           \@IEEEcompsocdiamondline\hfill\hbox{}\par}}}\relax
      \else
         \ifCLASSOPTIONtransmag
           {\@IEEEspecialpapernotice\mbox{}\vskip\@IEEEauthorblockconfadjspace%
            \mbox{}\hfill\begin{@IEEEauthorhalign}\@author\end{@IEEEauthorhalign}\hfill\mbox{}\par
           {\vspace{0.5\baselineskip}\relax\@IEEEtitleabstractindextextbox{\@IEEEtitleabstractindextext}\vspace{-1\baselineskip}\par}}\relax
         \else
           {\lineskip.5em\@IEEEcompsoconly{\sffamily}\sublargesize\@author\@IEEEspecialpapernotice\par
           {\@IEEEcompsoconly{\vskip 1.5em\relax
            \@IEEEtitleabstractindextextbox{\@IEEEtitleabstractindextext}\par\hfill
            \@IEEEcompsocdiamondline\hfill\hbox{}\par}}}\relax
         \fi
      \fi
   \fi
\fi\par\addvspace{0.0\baselineskip}\egroup}

\def\EuMWtitlesize{\@setfontsize{\EuMWtitlesize}{24}{24pt}}
\def\EuMWauthorsize{\@setfontsize{\EuMWauthorsize}{11}{11pt}}
\def\EuMWaffilsize{\@setfontsize{\EuMWaffilsize}{10}{10pt}}
\def\EuMWcaptionsize{\@setfontsize{\EuMWcaptionsize}{9}{10pt}}
\def\EuMWbibsize{\@setfontsize{\EuMWbibsize}{8}{10pt}}

\def\@IEEEauthorblockNstyle{\EuMWauthorsize\@IEEEcompsocnotconfonly{\sffamily}\@IEEEcompsocconfonly{\large}}
\def\@IEEEauthorblockAstyle{\EuMWaffilsize\@IEEEcompsocnotconfonly{\sffamily}\@IEEEcompsocconfonly{\itshape}\@IEEEcompsocconfonly{\large}}
\def\@IEEEauthordefaulttextstyle{\EuMWauthorsize\@IEEEcompsocnotconfonly{\sffamily}\sublargesize}

\def\thebibliography#1{\section*{\refname}%
    \addcontentsline{toc}{section}{\refname}%
    \EuMWbibsize\@IEEEcompsocconfonly{\small}\vskip 0.3\baselineskip plus 0.1\baselineskip minus 0.1\baselineskip
    \list{\@biblabel{\@arabic\c@enumiv}}%
    {\settowidth\labelwidth{\@biblabel{#1}}%
    \leftmargin\labelwidth
    \advance\leftmargin\labelsep\relax
    \itemsep \IEEEbibitemsep\relax
    \usecounter{enumiv}%
    \let\p@enumiv\@empty
    \renewcommand\theenumiv{\@arabic\c@enumiv}}%
    \let\@IEEElatexbibitem\bibitem%
    \def\bibitem{\@IEEEbibitemprefix\@IEEElatexbibitem}%
\def\newblock{\hskip .11em plus .33em minus .07em}%
\ifCLASSOPTIONtechnote\sloppy\clubpenalty4000\widowpenalty4000\interlinepenalty100%
\else\sloppy\clubpenalty4000\widowpenalty4000\interlinepenalty500\fi%
    \sfcode`\.=1000\relax}

\long\def\@makecaption#1#2{%
\ifx\@captype\@IEEEtablestring%
\par\@IEEEtabletopskipstrut
\else
\@IEEEfigurecaptionsepspace
\fi
\setbox\@tempboxa\hbox{\normalfont\footnotesize {#1.}\nobreakspace\nobreakspace #2}%
\ifdim \wd\@tempboxa >\hsize%
\setbox\@tempboxa\hbox{\normalfont\footnotesize {#1.}\nobreakspace\nobreakspace}%
\parbox[t]{\hsize}{\normalfont\footnotesize\noindent\unhbox\@tempboxa#2}%
\else
\ifCLASSOPTIONconference \hbox to\hsize{\normalfont\footnotesize\hfil\box\@tempboxa\hfil}%
\else \hbox to\hsize{\normalfont\footnotesize\box\@tempboxa\hfil}%
\fi\fi
\ifx\@captype\@IEEEtablestring%
\@IEEEtablecaptionsepspace
\else
\fi}

\newlength\tablecaptiontotableskip
\newlength\figuretocaptionskip
\def\@IEEEfigurecaptionsepspace{\vskip\figuretocaptionskip\relax}%
\def\@IEEEtablecaptionsepspace{\vskip\tablecaptiontotableskip\relax}%

\def\abstract{\normalfont%
\@IEEEabskeysecsize\bfseries\textit{\abstractname}\,\bfseries\textit{---}\,%
\@IEEEgobbleleadPARNLSP}%

\def\IEEEkeywords{\normalfont%
\@IEEEabskeysecsize\bfseries\textit{\IEEEkeywordsname}\,\bfseries\textit{---}\,%
\@IEEEgobbleleadPARNLSP}%
\def\endIEEEkeywords{\relax\vspace{0.67ex}%
\par\if@twocolumn\else\endquotation\fi%
\normalsize\normalfont}%

\DeclareRobustCommand*{\EuMWauthorrefmark}[1]{\raisebox{0pt}[0pt][0pt]{\textsuperscript{\footnotesize{#1}}}}%
\def\@IEEEauthorblockNtopspace{0ex}
\def\@IEEEauthorblockAtopspace{1mm}
\def\tablename{Table}
\def\thetable{\arabic{table}}
\def\IEEEkeywordsname{Keywords}
\def\subsubsection{\@startsection{subsubsection}{3}{\z@}{1.5ex plus 1.5ex minus 0.5ex}%
{0.7ex plus .5ex minus 0ex}{\normalfont\normalsize\itshape}}%
\newlength{\CPheadmatchindent}%
\def\@seccntformat#1{\hbox to\CPheadmatchindent{\csname the#1dis\endcsname}\hskip 0.1em \relax}
\IEEEilabelindentA \parindent
\IEEEilabelindent \IEEEilabelindentA
\IEEEelabelindent \parindent
\IEEEdlabelindent \parindent
\IEEElabelindent \parindent
\makeatother

\begin{document}
\raggedbottom
%
%
%
\title{Angle-Equivariant Convolutional Neural Networks for Interference Mitigation in Automotive Radar}
%
%
\author{%
\IEEEauthorblockN{%
Christian Oswald\EuMWauthorrefmark{\#1},
Mate Toth\EuMWauthorrefmark{\#2},
Paul Meissner\EuMWauthorrefmark{*3}, 
Franz Pernkopf\EuMWauthorrefmark{\#4}
}
\IEEEauthorblockA{%
\EuMWauthorrefmark{\#}Signal Processing and Speech Communication Laboratory, Graz University of Technology, Austria\\
\EuMWauthorrefmark{*}Infineon Technologies AG, Graz, Austria\\
\{\EuMWauthorrefmark{1}christian.oswald, \EuMWauthorrefmark{2}mate.a.toth, \EuMWauthorrefmark{4}pernkopf\}@tugraz.at, \EuMWauthorrefmark{3}paul.meissner@infineon.com\\
}
}
%
\maketitle
%
%
\begin{abstract}
In automotive applications, frequency modulated continuous wave (FMCW) radar is an established technology to determine the distance, velocity and angle of objects in the vicinity of the vehicle. The quality of predictions might be seriously impaired if mutual interference between radar sensors occurs. Previous work processes data from the entire receiver array in parallel to increase interference mitigation quality using neural networks (NNs). However, these architectures do not generalize well across different angles of arrival (AoAs) of interferences and objects.
In this paper we introduce fully convolutional neural network (CNN) with rank-three convolutions which is able to transfer learned patterns between different AoAs. Our proposed architecture outperforms previous work while having higher robustness and a lower number of trainable parameters. We evaluate our network on a diverse data set and demonstrate its angle equivariance. 
\end{abstract}
\begin{IEEEkeywords}
FMCW radar, convolutional neural networks, interference mitigation, angle-equivariance, deep learning, complex-valued processing
\end{IEEEkeywords}


\section{Introduction}
FMCW radar operates by continuously transmitting a frequency modulated signal, which is subsequently reflected off objects. Demodulating the received with the transmitted signal allows the sensor to determine the range and velocity of these objects. Furthermore, multiple receive antennas can be used to estimate the angle of objects. However, FMCW radar mutual interference may occur if an interferer radar emits a frequency that is sufficiently close to the ego radar's frequency. In that case, interference appears as short bursts in the sensor's output which can strongly deteriorate detection performance. A more detailed description of mutual interference in FMCW radar can be found in \cite{brooker2007mutual}.

Some methods for interference mitigation such as frequency hopping \cite{bechter2016bats} try to avoid the occurrence of interference all-together, while others aim to remove interference patterns from an already corrupted signal. Examples include setting corrupted samples to zero \cite{fischer2016untersuchungen}, nonlinear filtering across frequency ramps \cite{wagner2018threshold}, and iterative adaptive thresholding (IMAT) \cite{marvasti2012sparse}. In recent years, the problem of interference mitigation has also been tackled with machine learning. Recurrent neural network architectures such as gated recurrent units can be used on the time-domain signal \cite{mun2018deep}, \cite{mun2020automotive}, where the latter is augmented with self-attention blocks. Reference \cite{rock2020deep} applies a CNN to denoise the range-Doppler (RD) map of a single antenna, where they use two separate input and output channels for the RD-map's real and imaginary part. In \cite{rock2021resource} this architecture was further optimized towards resource-efficency and has subsequently been implemented on an FPGA \cite{hirschmugl2022fast}. Other possible architectures for processing in the RD-domain include convolutional autoencoders \cite{fuchs2020automotive}, \cite{de2020deep}. The fully convolutional architecture described in \cite{ristea2020fully} transforms its input given as a spectrogram into an interference mitigated range-profile. In \cite{chen2021dnn} a combination of an autoencoder with a traditional interference detection filter is proposed. A complex-valued CNN, which processes the entire receiver array simultaneously to improve interference mitigation quality is introduced in \cite{fuchs2021complex}. 

\section{Angle-Equivariant CNNs}
 \label{sec:angle-equivariant}
 
We first review the two-dimensional complex-valued convolutional neural network (CCNN-2D) introduced in \cite{fuchs2021complex} and then use it as baseline to introduce our model. 
CCNN-2D consists of layers of complex-valued convolution kernels, activations and batch-normalization, and is applied to interfered multi-antenna FMCW radar data. More specifically, the inputs for the network are the two-dimensional Fourier transforms of the antennas' signals, the so-called range-Doppler (RD) maps. Each of the $N_A$ antenna's RD-map is treated as an individual two-dimensional complex-valued input channel for the network, which means that convolutions are performed along the range and Doppler dimension of the input. The network is then trained to perform a regression from $N_A$ interfered to $N_A$ clean RD-maps in a supervised manner, which means that the inputs and targets are identical except for the presence of interferences. 

\begin{figure}
\centering
\includegraphics[clip,width=\columnwidth] {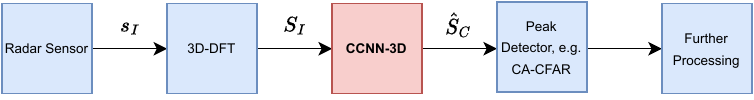}
\caption{Location of the proposed CCNN-3D in the digital signal processing chain. We denote the interfered multi-channel time domain signal as $s_I$. Its Fourier transform $S_I$ and the NN's prediction $\hat{S}_C$ are rank-three tensors with dimensions $[N_R, N_D, N_{\theta}]$, where $N_R$ is the number of samples per frequency modulation (FM) sweeps, $N_D$ the number of FM sweeps and $N_{\theta}$ the number of angle-bins.} \label{fig:processing_chain}
\vspace{-\baselineskip}
\end{figure}

\begin{figure}
\centering
\subfloat{%
  \includegraphics[clip,width=80mm]{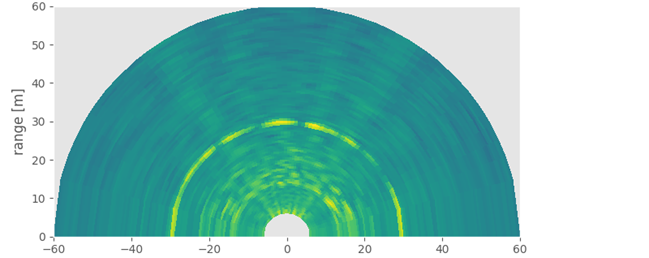}%
}

\subfloat{%
  \includegraphics[clip,width=80mm]{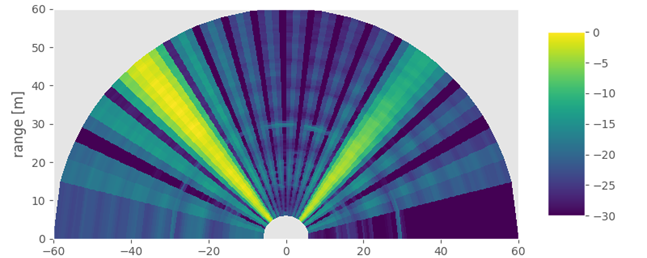}%
}

\subfloat{%
  \includegraphics[clip,width=80mm]{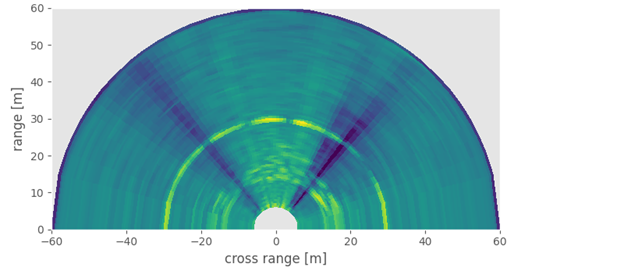}%
}

\caption{Exemplary data sample used to train the proposed CCNN-3D, depicted as range-angle maps. The \textbf{top map} depicts a clean sample, where object locations are visible as peaks. Clean samples are used as optimization targets during training. The \textbf{middle map} shows the same sample corrupted by an interference impending from roughly 45 degrees, masking the objects. Interfered samples are used as input for CCNN-3D. The \textbf{bottom map} shows the CCNN-3D's prediction for the middle map. We have up-sampled the plots' angle resolution for better interpretability; Note that no such up-sampling is needed when feeding RDA-maps to CCNN-3D. Each map is scaled such that its maximum value is zero dB.} \label{fig:range-azimuth}
\end{figure}

\begin{figure*}[t]
\centering
\includegraphics[width=140mm]{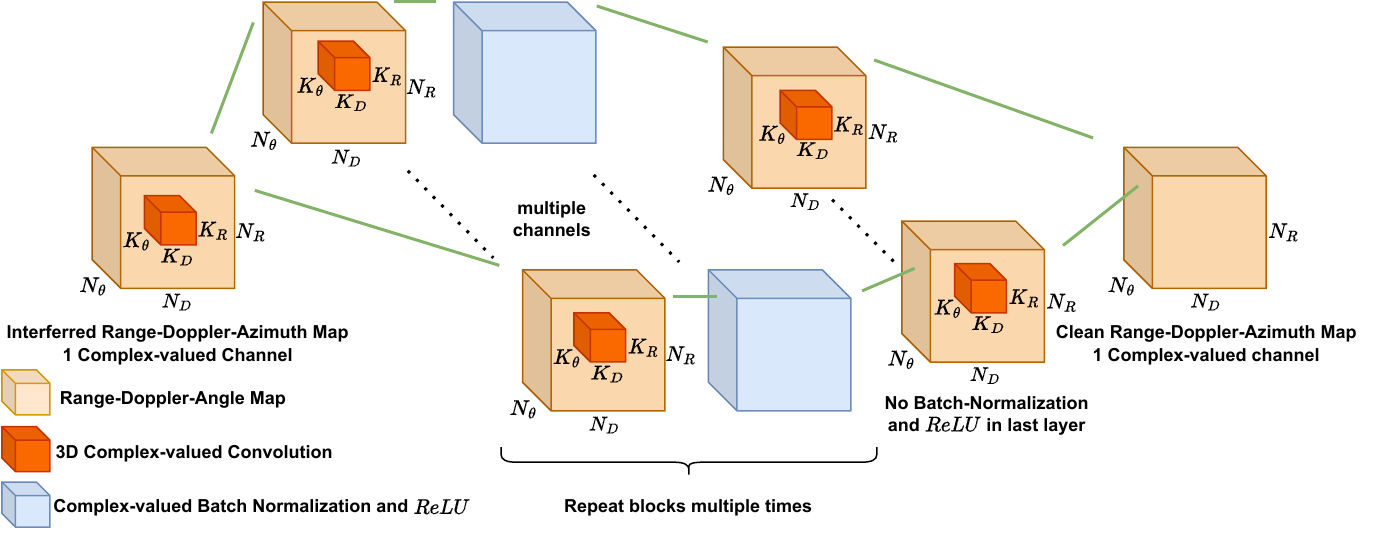}
\caption{Overview of the proposed structure. The network's input, output and activations consist of $N_R$ range, $N_D$ doppler and $N_\theta$ angle-bins, and convolution kernels have size $[K_R, K_D, K_{\theta}]$.} \label{fig:architecture}
\vspace{-\baselineskip}
\end{figure*}

The novel architecture presented in this paper labelled CCNN-3D is an extension of CCNN-2D and is depicted in \Fref{fig:architecture}. We replace the input/target tuples of multi-channel RD-maps $S_{RD}[r,d,a]$ by a single-channel range-Doppler-angle (RDA) map $S[r,d,\theta]$. An RDA-map is obtained by a third DFT over the antenna dimension of the multi-antenna RD-maps. An exemplary data sample used to train CCNN-3D can be seen in \Fref{fig:range-azimuth}, where we have performed a non-coherent summation over all Doppler bins. The placement of CCNN-3D in the overall processing chain is visualized in \Fref{fig:processing_chain}. The main advantage of representing the radar's signal as RDA-maps is that they capture the locality of interferences in the angle dimension. If the ego radar is interfered by another radar from an angle $\theta_0$, most of the interference's energy is located around the RDA-map's angle-bins corresponding to $\theta_0$ and $-\theta_0$ \cite{bechter_analytical_2017}. Furthermore, shifting an interference's AoA results in a shifted interference pattern in the RDA-map's angle dimension. These properties of locality and similarity can be leveraged by using convolutions in the angle dimension, resulting in rank-three convolution kernels. CCNN-3D does not use fully connected or pooling layers, i.e., shift-equivariance can be guaranteed in all three dimensions. Therefore, CCNN-3D is shift-equivariant w.r.t. the angle in addition to CCNN-2D's shift-equivariance w.r.t. to range and Doppler. In general, the convolution operator is shift-equivariant as $W * \nu(x) = \nu(W * x)$, where $x$ is a signal, $\nu$ is the shift operator and $W$ is the convolution kernel. CCNN-2D can be viewed as being fully-connected w.r.t. the antenna dimension; Replacing rank-two with rank-three convolutions therefore also strongly reduces the model's number of parameters. The convolution's stride is always one and the input to each convolution kernel is zero-padded such that the convolution output has the same size as the input. As all activations are now rank-three, we replace rank-two complex-valued $ReLU$ and batch-normalization by their rank-three variants. Note that the exact same architecture can be used for varying input sizes, as CCNN-3D operates independently from the input's number of range, Doppler and angle bins. If the used radar sensor measures both azimuth and elevation, all rank-three operations can be extended to their rank-four counterparts to perform interference mitigation on range-Doppler-azimuth-elevation maps in an analogous manner. 

\section{Experimental Setup}
In this section, we compare the proposed CCNN-3D to other interference mitigation methods, namely CCNN-2D \cite{fuchs2021complex}, zeroing \cite{fischer2016untersuchungen}, ramp-filtering \cite{wagner2018threshold} and IMAT \cite{marvasti2012sparse}.

\subsection{Data Set}
\label{sec:Dataset}
The targets of our data set consist of real-world inner-city measurements. We then generate and add artificial interferences to these targets, which are used as input to the CCNNs. We generate artificial interference FMCW signals by sampling uniformly from a range of radar parameters such as sweep duration (12 µs to 24 µs), sweep bandwidth (0,15 GHz to 0,25 GHz), AoA ($-90^{\circ}$ to $90^{\circ}$), sweep starting frequency (78,9 GHz to 79,1 GHz), number of sweeps (100 to 156), signal-plus-noise-to-interference ratio (30 dB to 50 dB), and number of interferers (1 to 3).
One sample consists of $96$ range, $96$ Doppler and $16$ angle bins, i.e., one sample is a rank-three tensor with dimensions $[96,96,16]$. The data is subsequently split into a training, test and validation set with size 2500, 250 and 250, respectively.

\subsection{Evaluation Metrics} \label{sec:metrics}
As FMCW radar sensors are primarily used for object detection, we run a cell averaging-constant false alarm rate (CA-CFAR) detector \cite{scharf1991statistical} on the network's output to verify that objects are no longer masked by interferences. We compute a non-coherent sum over the antenna dimension of a multi-antenna RD-map $S_{RD}[r,d,a]$ or equivalently over the angle dimension of an RDA-map $S[r,d,\theta]$ before feeding data into the CA-CFAR detector. 
The CA-CFAR detector therefore returns a rank-two binary object map. 
\subsubsection{F1-score}
The object map computed from the CCNN's output can be compared to the clean sample's object map using the F1-score, which is defined as 
\begin{equation}
    F_1 = 2 \cdot \frac{N_{TP}}{N_{TP} + \frac{1}{2} (N_{FP} + N_{FN})},
\end{equation}
where the number of true positives $N_{TP}$, false positives $N_{FP}$ and false negatives $N_{FN}$ are obtained by element-wise comparison of the object maps. For the computation of the F1-score we only keep the peaks of detected object clusters to account for extended objects.

\subsubsection{Error Vector Magnitude (EVM)}
We use the EVM to gauge the deviation of the predicted multi-antenna RD-maps $\hat{S}_{RD,C}$ from the ground-truth maps $S_{RD,C}$ at ground-truth object peak locations.
It can be computed as
\begin{equation}
    \text{EVM} = \frac{1}{N_A N_\mathcal{O}} \underset{a}{\sum} \underset{\{r,d\} \in \mathcal{O}}{\sum} \frac{| S_{RD,C}[r,d,a] -  \hat{S}_{RD,C}[r,d,a]|}{|S_{RD,C}[r,d,a]|} \,
\end{equation}
Where $N_A$ is the number of antennas $a$, $\mathcal{O}$ is the set of ground truth object peaks, given by their location coordinates $r$ and $d$ and $N_{\mathcal{O}}$ is the number of object peaks in an object map. 

\subsubsection{Peak Phase Mean Squared Error (PPMSE)}
The PPMSE is a measure of difference between the predicted and the ground-truth RD-maps' phase, evaluated at ground-truth object peak locations. It is given by
\begin{gather} 
    \Delta [r, d, a] = |\angle \hat{S}_{RD,C}[r,d,a] - \angle S_{RD,C}[r,d,a]|, \\
    \text{PPMSE} = \frac{1}{N_A N_{\mathcal{O}}} \underset{a}{\sum} \underset{\{r,d\} \in \mathcal{O}}
    {\sum} \text{min}(\Delta [r, d, a], 2 \pi - \Delta [r, d, a])^2.
\end{gather}
Note that we need to perform an inverse DFT on CCNN-3D's output to evaluate the EVM and the PPMSE, since CCNN-3D outputs RDA-maps. 
\subsection{Evaluated Architectures \& Training Setup}
We evaluate four instantiations of CCNN-3D which differ by the number and width of layers. All variants of CCNN-3D use kernels with size $[3,3,3]$. We also train one variant of CCNN-2D with kernel size $[3,3]$, which has been shown to perform well in experiments conducted in \cite{fuchs2021complex}. The evaluated CCNNs are summarized in \Fref{tab:archs}. We omit batch-normalization in the first layer on all CCNNs as it does not seem to improve performance.
{
%
\setlength{\tabcolsep}{2mm}%
\renewcommand{\arraystretch}{1.2}
\begin{table}[H]
\caption{Evaluated architectures. The number of channels of the $n$-th layer is given as the $n$-th element in the list.} \label{tab:archs}
\small
\centering
\begin{tabular}{|l|l|l|}\hline
\raisebox{-0.25mm}{Name} & \raisebox{-0.25mm}{Structure} & \raisebox{-0.25mm}{\#Parameters} \\ \hline \hline
CCNN-3D-l & $[32,16,8,4,1]$ & 38494 \\ \hline
CCNN-3D-m & $[16,8,4,2,1]$ & 10176 \\ \hline
CCNN-3D-s & $[8,4,2,1]$  & 2760 \\ \hline
CCNN-3D-xs & $[4,2,1]$ & 780 \\ \hline
CCNN-2D & $[32,16,16]$ & 23152 \\ \hline
\end{tabular}
\end{table}
}
We trained all CCNNs using a batch-size of 8 with a decaying learning rate for a maximum of 100 epochs. We use ADAM to minimize the mean squared error (MSE) and perform early stopping w.r.t. the validation MSE. To make results comparable, we do not perform windowing before the DFT over the antenna dimension when feeding data into CCNN-3D.

\section{Experiments}
\label{sec:Experiments}
We evaluate our proposed CCNN-3D on the data set described in \Fref{sec:Dataset} and compare it to CCNN-2D \cite{fuchs2021complex} as well as classical signal processing methods.
As shown in \Fref{tab:performance}, CCNN-3D leads to a higher F1-score than other methods, which indicates that its output is the most suitable for subsequent object detection. Interestingly, this performance gap vanishes when comparing the EVM and PPMSE. We hypothesize that CCNN also suppresses noise in addition to interferences and only keeps objects in its prediction, but further research is necessary to validate this interpretation. 

CCNN-3D-m and CCNN-3D-s only perform slightly worse than CCNN-3D-l while requiring a fraction of computational costs and trainable parameters. Even though CCNN-3D-xs has fewer parameters than CCNN-2D, memory requirements increase, as activations are rank-three instead of rank-two tensors. Consequently, the number of floating point operations required to process these intermediate values also grows. \\
{
%
\setlength{\tabcolsep}{2mm}%
\renewcommand{\arraystretch}{1.2}
\begin{table}[H]
\caption{Performance after training on the data set described in \Fref{sec:Dataset}.} \label{tab:performance}
\small
\centering
\begin{tabular}{|l||l|l|l|l|}\hline
\raisebox{-0.25mm}{Mitigation Technique} & \raisebox{-0.25mm}{F1} & \raisebox{-0.25mm}{EVM} & \raisebox{-0.25mm}{PPMSE} \\ \hline \hline
CCNN-3D-l \textbf{(ours)}  &  \textbf{0.9673}  & \textbf{0.5028} &  \textbf{0.3352} \\ \hline
CCNN-3D-m \textbf{(ours)}  &  0.9595  & 0.6153 &  0.4174 \\ \hline
CCNN-3D-s \textbf{(ours)}  &  0.9460 &  0.8232  &  0.5407 \\ \hline
CCNN-3D-xs \textbf{(ours)}  & 0.9151   & 0.9686  &   0.6464 \\ \hline
CCNN-2D \cite{fuchs2021complex}           & 0.8351  & 1.1766  &  0.7767 \\ \hline
Zeroing \cite{fischer2016untersuchungen}       & 0.7635   & 0.7121   &   0.5811  \\ \hline
Ramp-Filtering \cite{wagner2018threshold}  & 0.8403    & 0.5368     &   0.3679   \\ \hline
IMAT \cite{marvasti2012sparse}     & 0.7880    & 0.7280      &   0.5812  \\ \hline
No Mitigation & 0.5446 & 1.6740 & 0.6842 \\ \hline
\end{tabular}
\label{tab:fontsizes}
\end{table}
}

Furthermore, we compare CCNN's generalization capabilities w.r.t. the interferences' AoA. The training and validation data set used in this experiment are identical to the data set described in \Fref{sec:Dataset}, except that now we fix the interferences' AoA to 45°. We then report results on the same test set as above, i.e., all AoAs are allowed. 
{
%
\setlength{\tabcolsep}{2mm}%
\renewcommand{\arraystretch}{1.2}
\begin{table}[H]
\caption{Performance on test set with uniform AoA of interferences after training on data with a fixed AoA of 45°.} \label{tab:angle-equivariance} 
\small
\centering
\begin{tabular}{|l||l|l|l|l|}\hline
\raisebox{-0.25mm}{Mitigation Technique} & \raisebox{-0.25mm}{F1} & \raisebox{-0.25mm}{EVM} & \raisebox{-0.25mm}{PPMSE} \\ \hline \hline
CCNN-3D-l \textbf{(ours)}  &  0.8303  & 0.7922 &  \textbf{0.4702} \\ \hline
CCNN-3D-m \textbf{(ours)}  &  0.8897  & \textbf{0.7416} &  0.4808 \\ \hline
CCNN-3D-s \textbf{(ours)}  &  \textbf{0.9128}  &  0.8615  &  0.5545 \\ \hline
CCNN-3D-xs \textbf{(ours)}  &  0.8842  &  1.0167  &  0.6800 \\ \hline
CCNN-2D \cite{fuchs2021complex}      & 0.5349 & 1.7410 &  0.8064 \\ \hline
\end{tabular}
\end{table}
}
\Fref{tab:angle-equivariance} shows the improved generalization capabilities of CCNN-3D compared to CCNN-2D. Interestingly, smaller models now outperform CCNN-3D-l w.r.t. F1 and EVM, which indicates that CCNN-3D-l is overfitting to the interferences' fixed AoA in the training and validation set.

Compared to \Fref{tab:performance}, CCNN-3D's performance drops, as the appearance of an interference still changes when varying its AoA by a fraction of an angle-bin. For instance, when an interference's AoA perfectly coincides with an angle-bin, it will appear as a Kronecker-delta in the RDA-map. By contrast, when an interference's AoA is located between two neighbouring angle-bins, the corresponding interference pattern in the RDA-map will have much wider spread in the angle-dimension. We expect angle generalization to improve when using radar sensors with more receive antennas.
\section{Conclusion}
Our proposed architecture has fewer parameters and higher robustness compared to previous work as it generalizes across all AoAs. Nevertheless, classical signal processing methods for interference mitigation are more transparent and computationally cheaper.
In future work we therefore aim to reduce the model's computational footprint by considering the independence of range, Doppler and angle of objects. Furthermore, we plan to increase CCNN-3D's transparency by directly training it on object detections \cite{oswald_ete}.

\bibliographystyle{IEEEtran}

\bibliography{IEEEabrv,IEEEexample}

\begin{thebibliography}{10}
\providecommand{\url}[1]{#1}
\csname url@samestyle\endcsname
\providecommand{\newblock}{\relax}
\providecommand{\bibinfo}[2]{#2}
\providecommand{\BIBentrySTDinterwordspacing}{\spaceskip=0pt\relax}
\providecommand{\BIBentryALTinterwordstretchfactor}{4}
\providecommand{\BIBentryALTinterwordspacing}{\spaceskip=\fontdimen2\font plus
\BIBentryALTinterwordstretchfactor\fontdimen3\font minus \fontdimen4\font\relax}
\providecommand{\BIBforeignlanguage}[2]{{%
\expandafter\ifx\csname l@#1\endcsname\relax
\typeout{** WARNING: IEEEtran.bst: No hyphenation pattern has been}%
\typeout{** loaded for the language `#1'. Using the pattern for}%
\typeout{** the default language instead.}%
\else
\language=\csname l@#1\endcsname
\fi
#2}}
\providecommand{\BIBdecl}{\relax}
\BIBdecl

\bibitem{brooker2007mutual}
G.~M. Brooker, ``Mutual interference of millimeter-wave radar systems,'' \emph{IEEE Transactions on Electromagnetic Compatibility}, vol.~49, no.~1, pp. 170--181, 2007.

\bibitem{bechter2016bats}
J.~Bechter, C.~Sippel, and C.~Waldschmidt, ``Bats-inspired frequency hopping for mitigation of interference between automotive radars,'' in \emph{2016 IEEE MTT-S International Conference on Microwaves for Intelligent Mobility (ICMIM)}.\hskip 1em plus 0.5em minus 0.4em\relax IEEE, 2016, pp. 1--4.

\bibitem{fischer2016untersuchungen}
C.~Fischer, \emph{Untersuchungen zum interferenzverhalten automobiler radarsensorik}.\hskip 1em plus 0.5em minus 0.4em\relax Cuvillier Verlag, 2016.

\bibitem{wagner2018threshold}
M.~Wagner, F.~Sulejmani, A.~Melzer, P.~Meissner, and M.~Huemer, ``Threshold-free interference cancellation method for automotive fmcw radar systems,'' in \emph{2018 IEEE International Symposium on Circuits and Systems (ISCAS)}.\hskip 1em plus 0.5em minus 0.4em\relax IEEE, 2018, pp. 1--4.

\bibitem{marvasti2012sparse}
F.~Marvasti, M.~Azghani, P.~Imani, P.~Pakrouh, S.~J. Heydari, A.~Golmohammadi, A.~Kazerouni, and M.~Khalili, ``Sparse signal processing using iterative method with adaptive thresholding (imat),'' in \emph{2012 19th International Conference on Telecommunications (ICT)}.\hskip 1em plus 0.5em minus 0.4em\relax IEEE, 2012, pp. 1--6.

\bibitem{mun2018deep}
J.~Mun, H.~Kim, and J.~Lee, ``A deep learning approach for automotive radar interference mitigation,'' in \emph{2018 IEEE 88th Vehicular Technology Conference (VTC-Fall)}.\hskip 1em plus 0.5em minus 0.4em\relax IEEE, 2018, pp. 1--5.

\bibitem{mun2020automotive}
J.~Mun, S.~Ha, and J.~Lee, ``Automotive radar signal interference mitigation using rnn with self attention,'' in \emph{ICASSP 2020-2020 IEEE International Conference on Acoustics, Speech and Signal Processing (ICASSP)}.\hskip 1em plus 0.5em minus 0.4em\relax IEEE, 2020, pp. 3802--3806.

\bibitem{rock2020deep}
J.~Rock, M.~Toth, P.~Meissner, and F.~Pernkopf, ``Deep interference mitigation and denoising of real-world fmcw radar signals,'' in \emph{2020 IEEE International Radar Conference (RADAR)}.\hskip 1em plus 0.5em minus 0.4em\relax IEEE, 2020, pp. 624--629.

\bibitem{rock2021resource}
J.~Rock, W.~Roth, M.~Toth, P.~Meissner, and F.~Pernkopf, ``Resource-efficient deep neural networks for automotive radar interference mitigation,'' \emph{IEEE Journal of Selected Topics in Signal Processing}, vol.~15, no.~4, pp. 927--940, 2021.

\bibitem{hirschmugl2022fast}
M.~Hirschmugl, J.~Rock, P.~Meissner, and F.~Pernkopf, ``Fast and resource-efficient cnns for radar interference mitigation on embedded hardware,'' in \emph{2022 19th European Radar Conference (EuRAD)}.\hskip 1em plus 0.5em minus 0.4em\relax IEEE, 2022, pp. 1--4.

\bibitem{fuchs2020automotive}
J.~Fuchs, A.~Dubey, M.~L{\"u}bke, R.~Weigel, and F.~Lurz, ``Automotive radar interference mitigation using a convolutional autoencoder,'' in \emph{2020 IEEE International Radar Conference (RADAR)}.\hskip 1em plus 0.5em minus 0.4em\relax IEEE, 2020, pp. 315--320.

\bibitem{de2020deep}
M.~L.~L. de~Oliveira and M.~J. Bekooij, ``Deep convolutional autoencoder applied for noise reduction in range-doppler maps of fmcw radars,'' in \emph{2020 IEEE International Radar Conference (RADAR)}.\hskip 1em plus 0.5em minus 0.4em\relax IEEE, 2020, pp. 630--635.

\bibitem{ristea2020fully}
N.-C. Ristea, A.~Anghel, and R.~T. Ionescu, ``Fully convolutional neural networks for automotive radar interference mitigation,'' in \emph{2020 IEEE 92nd Vehicular Technology Conference (VTC2020-Fall)}.\hskip 1em plus 0.5em minus 0.4em\relax IEEE, 2020, pp. 1--5.

\bibitem{chen2021dnn}
S.~Chen, J.~Taghia, T.~Fei, U.~K{\"u}hnau, N.~Pohl, and R.~Martin, ``A dnn autoencoder for automotive radar interference mitigation,'' in \emph{ICASSP 2021-2021 IEEE International Conference on Acoustics, Speech and Signal Processing (ICASSP)}.\hskip 1em plus 0.5em minus 0.4em\relax IEEE, 2021, pp. 4065--4069.

\bibitem{fuchs2021complex}
A.~Fuchs, J.~Rock, M.~Toth, P.~Meissner, and F.~Pernkopf, ``Multi-antenna radar signal interference mitigation using complex-valued convolutional neural networks,'' Graz University of Technology, Tech. Rep., 2023.

\bibitem{bechter_analytical_2017}
J.~Bechter, M.~Rameez, and C.~Waldschmidt, ``Analytical and experimental investigations on mitigation of interference in a {DBF} {MIMO} radar,'' \emph{IEEE Trans. Microw. Theory Tech.}, vol.~65, no.~5, pp. 1727--1734, 2017.

\bibitem{scharf1991statistical}
L.~L. Scharf and C.~Demeure, \emph{Statistical signal processing: detection, estimation, and time series analysis}.\hskip 1em plus 0.5em minus 0.4em\relax Prentice Hall, 1991.

\bibitem{oswald_ete}
C.~Oswald, M.~Toth, P.~Meissner, and F.~Pernkopf, ``End-to-end training of neural networks for automotive radar interference mitigation,'' \emph{unpublished}, 2023.

\end{thebibliography}

\end{document}